**Investigating Dissemination of Scientific Information on Twitter: A Study of Topic Networks in Opioid Publications**

Robin Haunschild[1], Lutz Bornmann[2], Devendra Potnis[3], Iman Tahamtan[4, 5]

1. Max Planck Institute for Solid State Research, Stuttgart, Germany. Email address: r.haunschild@fkf.mpg.de

2. Division for Science and Innovation Studies, Administrative Headquarters of the Max Planck Society, Munich, Germany. Email address: bornmann@gv.mpg.de

3. School of Information Sciences, College of Communication and Information, University of Tennessee, Knoxville, TN, USA. Email address: dpotnis@utk.edu

4. School of Information Sciences, Communication, and Information, University of Tennessee, Knoxville, TN, USA. Email address: tahamtan@vols.utk.edu (Corresponding author)

5. The Machine Learning and Intelligence Operations Product Team, JP Morgan Chase, Columbus, Ohio.

## Abstract

One way to assess a certain aspect of the value of scientific research is to measure the attention it receives on social media. While previous research has mostly focused on the "number of mentions" of scientific research on social media, the current study applies "topic networks" to measure public attention to scientific research on Twitter. Topic networks are the networks of co-occurring author keywords in scholarly publications and networks of co-occurring hashtags in the tweets mentioning those scholarly publications. This study investigates which topics in opioid scholarly publications have received public attention on Twitter. Additionally, it investigates whether the topic networks generated from the publications tweeted by all accounts (bot and non-bot accounts) differ from those generated by non-bot accounts. Our analysis is based on a set of opioid scholarly publications from 2011 to 2019 and the tweets associated with them. We use co-occurrence network analysis to generate topic networks. Results indicated that Twitter users have mostly used generic terms to discuss opioid publications, such as "Opioid," "Pain," "Addiction," "Treatment," "Analgesics," "Abuse," "Overdose," and "Disorders." Results confirm that topic networks provide a legitimate method to visualize public discussions of health-related scholarly publications and how the Twitter users discuss health-related scientific research differently from the scientific community. There was a substantial overlap between the topic networks based on the tweets by all accounts and non-bot accounts. This result indicates that it might not be necessary to exclude bot accounts for generating topic networks as they have a negligible impact on the results. This study provided some preliminary evidence that scholarly publications have a network agenda-setting effect on Twitter, in that the networks of topics in scholarly publications can impact public attention on Twitter.

## 1. Introduction

### 1.1. Social media mentions in research evaluation

Social media increasingly play an important role in the dissemination of scientific information to the public. The public can then engage in discussions around scientific topics shared on social media (Chan et al., 2020; Murphy & Salomone, 2013; Patel et al., 2020). The transfer of scientific information to the public is an ongoing activity in which knowledge is obtained from those who own it (e.g., the authors and journals), is learned by those who don't have it (e.g., social media users), and is being passed to other people through social networks (Havakhor et al., 2018). People use social media to share and discuss complex scientific information (Murphy & Salomone, 2013), motivating public conversations around various issues such as climate change. The diffusion of scholarly publications and scientific information on social media can have a positive societal impact, for instance, by educating and promoting health literacy and changing public behavior (Korda & Itani, 2013).

To measure the impact of scientific information diffused on social media, most past studies have focused on the number of times scholarly publications have been mentioned on social media. The "number of mentions" of scholarly publications on social media has been used as a measure to evaluate the impact and value of scientific research (for a review see Bornmann, 2014; Sugimoto et al., 2017; Tahamtan & Bornmann, 2020). However, the number of mentions of scholarly publications on social media may only reflect public interest and attention to scientific research rather than their impact (Tahamtan & Bornmann, 2020).

## 1.2. Network of topics in scholarly publications in research evaluation

Haunschild, Leydesdorff and Bornmann (2020) and Haunschild, Leydesdorff, Bornmann, et al. (2019) suggested that besides "number of mentions", the impact of scientific information can also be evaluated by determining which topics in scholarly publications are more frequently discussed on social media. They noted that determining the topics in scholarly publications that have received public attention on social media provides a reasonable way to evaluate their impact beyond their academic impact that is often measured by citation counts (Haunschild, Leydesdorff, & Bornmann, 2020; Haunschild, Leydesdorff, Bornmann, et al., 2019).

Haunschild, Leydesdorff, Bornmann, et al. (2019) and Haunschild, Leydesdorff, Bornmann, et al. (2019) proposed a new method that not only measures public attention to scholarly publications on social media but also demonstrates how Twitter users (representing the public) discuss scientific research. They used a network approach in which a topic network (co-occurrence network of author keywords) in scholarly publications would be compared with a topic network of author keywords in scholarly publications that are mentioned on Twitter (or any other social media platform). This approach assumes that the topics in scholarly publications with broader societal impact would receive greater public attention on Twitter. The co-occurrence network approach focuses on the topics in a scholarly publication shared on Twitter rather than "number of mentions" or how many times the publication is mentioned on social media. An advantage of the co-occurrence network-based approach over previous approaches (measuring number of mentions) is that it can be used to analyze, map, and compare scientific discussions around a given topic (represented in the author keywords in those publications) with public discussions around that topic (assessed by the author keywords in the publications mentioned on Twitter).

### 1.3. Opioid scholarly publications on social media

The public has no access to scholarly publications about opioids (or any other topic) unless they are shared on social media or other platforms. That scholarly publications and the knowledge associated with them once shared on social media, would be opened to the public and create value for them.

The influence of opioid scholarly publications on public attention on social media and their role in creating public awareness is an understudied topic. Thus, it is important to study how scientific knowledge shared on social media via opioid scholarly publications creates value for users. In this study, we assess the topics in opioid scholarly publications that have received public attention on social media. The reason for studying opioids is declaring it as a public health emergency in the US in 2017 (The White House, 2017). Opioids cause the death of thousands of people worldwide every year (Rudd, 2016; Rudd et al., 2016). It is an important issue for the public which has many implications on areas such as public health, mental health, and economics. The results can be used to learn which topics in opioid scholarly publications have more impact or are more popular on social media. Therefore, it is of merit to study how opioid scholarly publications are discussed on Twitter.

### 1.4.  The knowledge gap and study objectives

As mentioned there has been considerable research on using numbers of mentions of scholarly publications on social media to evaluate the impact of research. However, little is known about using topic networks in research impact and assessing public attention to scholarly publications. Since one way to evaluate certain aspects of the value of scientific research is to measure the public attention it receives on social media and given the advantages of topic networks for assessing public attention to scientific research (Haunschild, Leydesdorff, & Bornmann, 2020; Haunschild, Leydesdorff, Bornmann, et al., 2019), this study uses the co-occurrence network analysis approach to assess which topics in "opioid" scholarly publications have received public attention on Twitter.

We also study the influence of bots in measuring public attention to scholarly publications on social media. A considerable amount of tweets are produced by automated social media accounts, known as bots (Ferrara, 2020b; Hegelich & Janetzko, 2016). Bots can impact public opinion and social media discussions by presenting a distorted reality, or artificially and forcefully changing or influencing the public discourse. Bots can manipulate public attention to and discussions on critical public issues such as COVID-19 (Ferrara, 2020a). Therefore, it is important to know if bots influence public attention to scholarly publications on social media. The influence of bots has not been investigated in previous Twitter network studies (Haunschild, Leydesdorff, & Bornmann, 2020; Haunschild, Leydesdorff, Bornmann, et al., 2019).

To address our research questions, we created co-occurrence networks of the author keywords in opioid scholarly publications from 2011-2019 in the Web of Science (WoS, Clarivate Analytics, Philadelphia, Pennsylvania, USA). We analyzed the topics associated with opioid scholarly publications shared on Twitter in comparison to the topics of opioid scholarly publications. The topic networks are shown in two different versions: (i) networks created by including the publications shared by all Twitter accounts (bot accounts and non-bot accounts) and (ii) networks created by including the publications shared by only non-bot Twitter accounts.

## 2. Background

### 2.1. Application of topic networks of scholarly publications shared on social media in research evaluation

Network analysis approaches can be used to analyze the diffusion of scholarly publications on social media. Alperin et al. (2019) studied the diffusion patterns of peer-reviewed scholarly publications on Twitter. They analyzed 1,590 tweets mentioning 11 articles in biology. Alperin et al. (2019) found that the users were connected through common followers that mostly shared the same article: most scholarly publications are spread on Twitter within a tightly connected single community. Additionally, almost half of the tweeted publications were disseminated on Twitter through retweets. Hellsten and Leydesdorff (2020) used a new network analysis approach to analyze the discussions around scholarly publications on Twitter for understanding and visualizing online science-related communications. Their approach indicated which Twitter users were connected with which hashtags.

Scholars such as Haunschild, Leydesdorff, Bornmann, et al. (2019), Haunschild, Leydesdorff, Bornmann, et al. (2019), and Haunschild, Leydesdorff, Bornmann, et al. (2020) proposed a network-based approach to measure public attention to scientific research on Twitter. Their network analysis approach can be used to identify (a) which scholarly publications have entered the public discussion on social media, (b) which topics in scholarly publications have received greater public attention on social media, and (c) how the public perceives and discusses scholarly publications differently from the scientific community (Haunschild, Leydesdorff, & Bornmann, 2020; Haunschild, Leydesdorff, Bornmann, et al., 2019).

Haunschild, Leydesdorff, Bornmann, et al. (2020) compared the topic network (network of author keywords) generated from approximately 46,000 climate change scholarly publications between 2011–2017 with the topic network in 775,499 tweets containing a link to those publications. They found that the climate change research topics that had achieved public attention on Twitter were generally related to the consequences of climate change for humans. They reported that publications with more general keywords were more likely to be tweeted than those with scientific jargon. In a similar study, using another topic, Haunschild, Leydesdorff and Bornmann (2020) examined how Library and Information Science (LIS) publications were discussed on Twitter. Their results demonstrated that only certain topics in LIS publications received public attention on Twitter, such as librarians, libraries, research, and social media. Haunschild, Leydesdorff and Bornmann (2020) also indicated that while all LIS publications were generally more focused on theoretical applications and methodologies, the topics in the tweeted LIS publications (and publications mentioned in the news) were related to health-applications, social media, privacy issues, and sociological studies.

The studies that have employed network analysis approaches (e.g., Haunschild, Leydesdorff, & Bornmann, 2020; Haunschild, Leydesdorff, Bornmann, et al., 2019) indicate that the dissemination of scholarly publications on social media brings public attention to some topics of scholarly outputs more than others. These studies also indicate that topics in scholarly publications are transferred to the public discussions as a bundle of networked topics, which is discussed in the following section.

## 2.2. Network agenda-setting effect of scholarly publications shared on social media

Scholarly publications shared on social media can set an agenda for social media users, consequently impacting, and shaping their opinion. We adopt this idea from the network agenda-setting model (Guo & McCombs, 2011) which states the repetition of an issue (e.g., opioid) and topics related to that issue (e.g., addiction, abuse, pain) will be transferred to and impact public opinion as a bundle of networked topics (Guo & McCombs, 2011).

According to the network agenda-setting model, it can be argued that the scholarly publications around any topic like opioids shared on social media, specifically by influential authors and high-impact journals, can set a public agenda, consequently attracting public attention and influencing the public opinion. For instance, Haunschild, Leydesdorff, Bornmann, et al. (2019) demonstrated that the terms related to climate policy such as "food security," "governance," "renewable energy," and "redd" (reducing emissions from deforestation and forest degradation) were linked to each other in a cluster of networked topics on Twitter. The topics adjacent to each other demonstrate that people would link and make connections between those topics in their minds, consequently impacting their opinion (Guo, 2015; Guo & McCombs, 2011).

To assess the network agenda-setting effect of scholarly publications on social media, the overlap (and/or correlation) between two topic networks can be evaluated: a topic network of author keywords in scholarly publications and a topic network of scholarly publications shared on social media and/or mentioned in the news. When there is a high overlap or correlation between the two networks, it can be said the agenda set by those scholarly publications have impacted public attention (see Guo, 2015; Guo & McCombs, 2011).

### 2.3. Bot accounts activity on social media

Bots can perform many actions such as information collection and distribution, generating clicks and contents (e.g., comments), and editing articles on Wikipedia (Leidl, 2019). Bots may intervene in online discussions on critical public issues to manipulate public opinion (Ferrara, 2020a). Some bots behave like humans (social bots) and are hard to detect (Hegelich & Janetzko, 2016). Previous studies have indicated that bots can impact public discussion on social media. For example, Hegelich and Janetzko (2016) showed that bots shape the public agenda on political issues. They noted that bots conceal their bot identity, prompt topics and hashtags to appeal to the public, and retweet selected tweets. For instance, bots take a popular tweet and retweet it by adding specific hashtags. Bots can prompt (political) topics and hashtags that may sound interesting to the public (Hegelich & Janetzko, 2016).

Only a few studies have investigated how bots impact scientific information diffusion on social media. Haustein et al. (2016) studied bot accounts that tweeted scholarly publications deposited on the preprint repository arXiv in 2012. They showed that bots created 9% of tweets to scholarly publications submitted to arXiv and were subsequently published in journals indexed in WoS. Bots distributing scholarly publications undermine the usefulness of tweets-based metrics for research impact (Haustein et al., 2016). Didegah et al. (2018) studied the impact of bots in distributing scholarly publications in five different fields. They found that 65% of Twitter accounts were bots that contributed to disseminating scientific information, particularly in life & earth science.

## 3. Data and Methods

### 3.1. Inclusion criteria for including studies

In the first step, we collected all the scholarly publications indexed in the WoS, published between 2011 and 2019, and contained opioid-related terms in their titles (see Table 1). The period of 2011 to 2019 was chosen because Altmetric.com started covering Twitter data in 2011 (Haunschild, Leydesdorff, Bornmann, et al., 2019). To retrieve publications from WoS, we needed a list of keywords relevant to opioids to be searched in WoS. The following section explains how the list of keywords was determined.

### 3.2. Methods to identify search terms

To find a list of keywords synonymous to opioid, we performed the following steps. The list of keywords is presented in Table 1.

First, we searched the Cochrane Database of Systematic Reviews (in the Cochrane Library, https://www.cochranelibrary.com/) in February 2020 to find the reviews that contained opioids or opioids in their titles. Fifty-seven reviews were retrieved. Cochrane reviews meticulously list the keywords used in their search strategy. We collected a list of opioid-related terms and synonyms from the 'Search Strategy' section of Cochrane reviews. Among the 57 retrieved reviews, we extracted the synonyms from four reviews published in the last two years. The most recent reviews dated back to 2019 (these studies include Candy et al., 2018; Doleman et al., 2018; Moe-Byrne et al., 2018; Smith et al., 2018). This approach provided us with a rich list of opioid synonyms such as Narcotics, Opiate, Morphine, Diamorphine (see Table 1).

Second, we found other relevant keywords from the following resources: the Centers for Disease Control and Prevention (CDC) annual surveillance report of drug-related risks and outcomes (Centers for Disease Control and Prevention, 2018a, 2019a), CDC guideline for prescribing opioids for chronic pain (Centers for Disease Control and Prevention, 2019c), and the National Institute on Drug Abuse (2019), and Rubinstein and Carpenter (2017).

Additionally, two public health experts (with research background on opioids) suggested "Suboxone," "Subutex," and "Heroin" to be added to the list.

Table 1. The list of sources that were used to determine the keywords used to perform the search in WoS

| Source | Keywords |
| --- | --- |
| Cochrane review: Candy et al. (2018); Doleman et al. (2018); Moe-Byrne et al. (2018); and Smith et al. (2018) | Narcotic, Opiate, Morphine, Diamorphine, Fentanyl, Remifentanil, Alfentanil, Meperidine, Pethidine, Tramadol, and Ketobemidone |
| Centers for Disease Control and Prevention (2018b); and Centers for Disease Control and Prevention (2019b) | Prescription pain relievers included opioids and covered the following drug subcategories: Hydrocodone, Oxycodone, Tramadol, Codeine, Morphine, Fentanyl, Buprenorphine, Butrans, Belbuca, Oxymorphone, Hydromorphone, Methadone, Tapentadol, and Propoxyphene |

| National Institute on Drug Abuse (2019) | Common prescription opioids: Hydrocodone (Vicodin), Oxycodone (OxyContin, Percocet), Oxymorphone (Opana), Morphine (Kadian, Avinza), Codeine, and Fentanyl<br><br>Slang terms used for opioid: Oxy, Percs, and Vikes |
|---|---|
| Rubinstein and Carpenter (2017) | Commonly prescribed opioids: Codeine, Fentanyl, Hydrocodone, Hydromorphone, Methadone, Morphine, and Oxycodone |
| Public Health Experts | Suboxone, Subutex, and Heroin |

### 3.3. Bibliometric data sources

The bibliometric data was collected from an in-house database developed and maintained by the Competence Centre for Bibliometrics (CCB, see: http://www.bibliometrie.info/) and retrieved from the Science Citation Index Expanded (SCI-EXPANDED), Social Sciences Citation Index (SSCI), and Arts and Humanities Citation Index (A&HCI), produced by Clarivate Analytics. The in-house database was last updated in April 2019. The search was performed in the WoS online interface. The export of the results was done using the "Fast 5K" mode. Only a few meta-data can be retrieved this way, e.g., author keywords are excluded. Thus, we extracted the WoS UTs (a unique accession number of a record in the WoS) and DOIs and appended the author keywords from the in-house database.

In the first step, we searched a combination of search terms mentioned in Table 1 in the title (TI) field of documents. Our initial search indicated that some of the retrieved documents were not related to opioids. For example, a document that included "Oxy-fuel" in its title was assessed as irrelevant. Besides, some search terms such as "Percs" only retrieved two documents, and "Vikes" did not retrieve any documents. These search terms, which are the slang used for opioids (National Institute on Drug Abuse, 2019) were excluded from our search. We performed our search, including the final list of search terms in WoS, in three steps (see Appendix A).

The publication year 2019 was incomplete at the time of data retrieval (February 2020). However, this is irrelevant as our in-house database was last updated at the end of April 2019. We were able to match 14,381 UTs in the in-house database. For all practical purposes, we expect all opioid publications between 2011 and 2018, with a few early indexed opioid publications from 2019 to be in our dataset. Of those 14,381 publications, 10,855 contained author keywords.

### 3.1. Data extraction from Twitter and News outlets

Altmetric.com (see https://www.altmetric.com) is a company that tracks mentions of scholarly publications in various sources such as Twitter, Facebook, news outlets, and Wikipedia. Scholarly publications' mentions can be accessed at no cost for research purposes via the Altmetric.com API or snapshots. Altmetric.com provides access to the IDs of tweets (a unique identification number assigned to each tweet by Twitter). These IDs were used to download 173,187 tweets (including retweets) associated with 6,433 opioid publications tweeted by at least two different accounts via the Twitter API. We did not include publications tweeted only once to reduce noise in the data because we assume they may have been tweeted by the publisher or the authors for self-promotion purposes. We also downloaded other available information besides the tweet texts from the Twitter API using the R software (R Core Team, 2019) such as Twitter user names (see Appendix B). The number of times a paper was mentioned in tweets or news outlets was taken from the Altmetric.com database snapshot.

Some tweets were not available and accessible such as "private tweets," "deleted tweets," and "suspended accounts," therefore were not included in our analysis (see Appendix C).

The analysis units in this study were author keywords in opioid publications and hashtags associated with the tweets mentioning opioid publications (as the representation of topics). Three sets of author keywords were extracted: (1) author keywords in all opioid publications, (2) author keywords in publications tweeted at least twice, and (3) author keywords of publications tweeted at least twice and mentioned in the news outlets. As mentioned, publications tweeted only once were excluded because they may have been tweeted by the publisher or the authors for self-promotion purposes.

### 3.2. Detection of bots

Detecting bots is very hard due to the evolving capabilities of bots. However, many attempts have been made to detect bots in recent years (Ferrara, 2020b). This study used the default model of the R package "tweetbotornot" designed by Kearney (2019) to detect bots. The default model of "tweetbotornot" is 93.53% accurate in classifying bots and 95.32% accurate in classifying non-bots (Kearney, 2019). It uses two sets of data to classify accounts to bots and non-bots: (i) users-level data such as biography, location, number of followers and friends, and (ii) tweets-level data such as the number of hashtags, mentions, and capital letters in the 100 most recent tweets of a Twitter user (Kearney, 2019). This model can only classify 180 users as bots or non-bots every 15 minutes. Accounts that receive a score of at least 0.5 (a probability of 50%) are considered bots (Davis et al., 2016).

We ran the "tweetbotornot" package on the Twitter accounts (usernames) in our dataset (56,266 distinct users), which took 3.256 days (180 users per 15 minutes) to classify Twitter accounts to bots and non-bots. Running the package on our dataset, we obtained two warning messages for some accounts: "sorry, that page does not exist" and "not authorized." We used a self-consistent methodology to find the bot probability of the profiles that returned a 'not authorized' error (n=22,396). We re-ran the package on these 22,396 profiles. This resulted in identifying a bot probability for 21,416 of the accounts and an error for 980. We re-ran the package on the 980 accounts and received a valid bot estimate for 188 accounts and an error for 792. Re-running the package on these 792 accounts resulted in errors for all. Thus, we stopped re-running the package on these accounts. Overall, we found the bot probability of up to 50% for 28,985 accounts (non-bots) and above 50% for 26,489 accounts (bots). The following histogram shows the plot of the bot probability scores with a red line at 0.5.

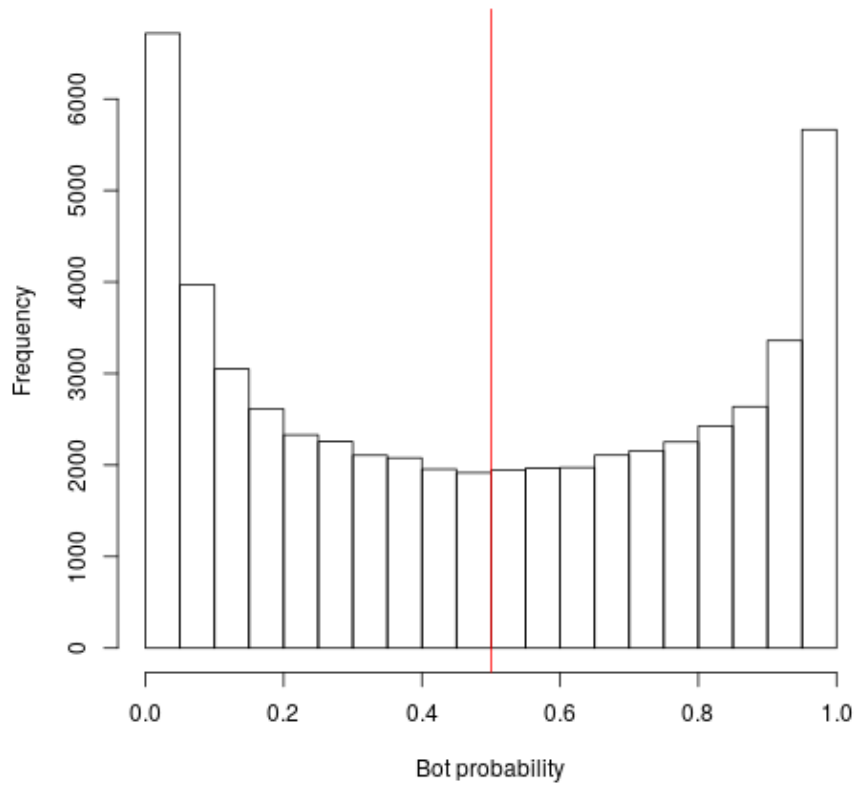

Figure 1. The plot of bot probability scores for Twitter accounts discussing the opioid scholarly publications published between 2011 and 2019 and indexed in WoS. Accounts with a probability score up to 0.5 are considered as non-bots (n=28,985), and accounts with a score above 0.5 are considered bots (n=26,489).

### 3.1. Visualization of networks

To visualize the data, we used VOSviewer v.1.6.12 (http://www.vosviewer.com). Following Haunschild, Leydesdorff and Bornmann (2019, 2020), the algorithm designed by Kamada and Kawai (1989) for drawing undirected graphs and weighted graphs was used to lay out the resulting files (containing cosine-normalized distributions of terms in the Pajek format, see http://mrvar.fdv.uni-lj.si/pajek). The community-searching algorithm in VOSviewer was employed with a resolution parameter of 1.0, the minimum cluster size of 1, 10 random starts, ten iterations, a random seed of 0, and the option 'merge small clusters' enabled. The node's size indicates the frequency of co-occurrence of a specific term with all other terms on the network. Lines between two nodes and their thickness indicate the co-occurrence frequency of these particular terms.

Before creating the networks, we unified some synonyms in our dataset using Excel. For example, "Drug_abuse," "Substance_abuse," and "Opioid_abuse" were merged into "Opioid_abuse." However, we did not combine general terms (i.e., "Abuse," "Treatment," "Prescribing," "Addiction," "Dependence," "Overdose," "Analgesic," "Constipation," and "Hyperalgesia") to the categories with more specific terms. For example, we considered "Prescribing" as a general term, but more specific terms such as "Opioid_prescribing," "Prescription_drugs," and "Prescription_opioid" were merged into "Prescription_opioid". In addition, "Opioids," "Opiates," "Opiate," and "Opioid" were all grouped as 'Opioid'. We also did the same grouping for other general terms such as "Analgesia" and "Analgesics" grouped as "Analgesics," or "Narcotics" and "Narcotic" which were grouped as "Narcotic."

We only included the most frequently occurring author keywords and hashtags in our analysis: We used the author keywords that appeared more than five times in opioid publications and were tweeted by at least two accounts and were mentioned at least once in news outlets. The resulting number of top author keywords was 70. The other sets of publications and their author keywords were larger. To compare networks and data of similar sizes, we also tried to use the top-70 author

keywords for the other sets (all publications and tweeted publications). However, due to the tied author keywords, a slightly different number of most frequent author keywords had to be used in some cases, e.g., top-69 author keywords for the publications that were tweeted by at least two accounts.

One network from the top-70 author keywords in all opioid publications were created (see Figure 2 below). Also, five networks from the tweets that were posted by all accounts (bot and non-bot accounts) were created: one network with the top-69 author keywords of opioid publications that were tweeted by at least two accounts (see Figure 3 below), one network with the top-64 author keywords of publications tweeted by at least two accounts and mentioned in the news (see Figure 5 below), one network with the top-70 hashtags of tweets (see Figure 7 below), and one network of the top-35 author keywords tweeted by at least two accounts and top-35 hashtags of tweets (see Figure 9 below).

We also created five networks from the tweets that were posted by only non-bot accounts: One network with the top-64 author keywords of opioid publications that were tweeted by at least two accounts (see Figure 4), one network with the top-64 author keywords of publications tweeted by at least two accounts and mentioned in the news (see Figure 6), one network with the top-64 hashtags of tweets (see Figure 8), and one network of the top-35 author keywords tweeted by at least two accounts and top-35 hashtags of tweets (see Figure 10, a network of 70 hashtags/keywords).

## 4. Results

The following section presents various networks used to explore the similarities and differences of scientific and public communications around opioids on Twitter.

### 4.1. Author keywords

Figure 2 shows a network of the top-70 author keywords in the opioid scholarly publications published between 2011 and 2019. The six clusters of author keywords marked by respective colors are visualized in the figure, which provides a general overview of the topics presented in opioid scholarly publications from 2011 to 2019. Each cluster includes a set of related keywords in each network.

The largest node in Figure 2 is the term 'Opioid' in the center of the blue network, surrounded by terms that have close ties with opioids such as opioid disorders, opioid addiction, opioid treatment, opioid use. The green cluster contains pain-related terms including pain management, neuropathic pain, postoperative pain, acute pain, and analgesic. The red cluster comprises terms related to the use and dependence on substances like morphine, dynorphin, dopamine, and cocaine which are also used as pain relievers. This cluster also contains mental disorders and issues related to the consumption of opioids such as anxiety, depression, and stress. The light blue cluster deals with opioid prescriptions, use, abuse, and overdose. The light blue and purple clusters comprise author keywords related to the consequences of taking opioids for a long time, mainly constipation and abnormally increased sensitivity to pain (hyperalgesia).

Figure 2. Network of top-70 author keywords in all opioid publications published between 2011 and 2019. An interactive version of this network can be viewed at https://s.gwdg.de/gWGMIa.

The network of opioid publications tweeted by at least two accounts is illustrated in Figure 3 (bot and non-bot) and Figure 4 (only non-bot accounts). There was a high overlap (95.3%) between the keywords in these two networks, which might indicate that bots do not impact public attention to opioid scholarly publications on Twitter or that bot accounts communicate scholarly publications similar to humans.

The green cluster in Figure 3 consists of terms related to opioid misuse, overdose, analgesics, epidemic, prescription, and primary care. The red cluster includes author keywords that deal with opioid use, addiction, disorders, and treatment. The blue cluster contains keywords related to pain management and palliative care. The yellow cluster is pertinent to the use of post-operative analgesic drugs. Most of the keywords in Figure 3 also appear in Figure 4, and the two networks have 61 keywords in common (95.3%).

Figure 3. Network of top-69 author keywords in opioid publications published between 2011 and 2019, tweeted by at least two accounts (considering all accounts). The interactive network can be found at: https://s.gwdg.de/EhmvtM.

Figure 4. Network of top-64 author keywords in opioid publications published between 2011 and 2019, tweeted by at least two accounts (considering only non-bot accounts). The interactive network can be found at: https://s.gwdg.de/cRGyNN.

Figure 5 shows the top-64 author keywords in publications tweeted by at least two accounts (considering all accounts) and mentioned in the news. Figure 6 shows the top-64 author keywords in publications tweeted by at least two accounts (considering only non-bot accounts) and mentioned in the news. We tried to focus more specifically on the public discourse by including the news outlets because we expect the news editors to select topics that are most certainly of public interest.

In Figure 5, the red cluster is centered around pain management and primary and palliative care. The green cluster consists of opioid misuse, anxiety, and depression. The blue color is focused on self-administration and dependence on alcohol, cocaine, codeine, tramadol, and heroin. The yellow cluster is comprised of opioid use, addiction, and treatment. The purple cluster seems pertinent to making opioid-related policies such as policies regarding opioid prescription, and policies about the use of marijuana and cannabis. 100% (n=64) of the author keywords in Figure 5 (considering all accounts) also appear in Figure 6 (considering only non-bot accounts), yet they may appear in different clusters (with different colors). This high overlap further provides evidence that bots may not impact topic networks of opioid scholarly publications.

Figure 5. Network of top-64 author keywords in publications published between 2011 and 2019, tweeted by at least two accounts (considering all accounts) and mentioned in the news. The interactive network can be found at: https://s.gwdg.de/YfAXwY.

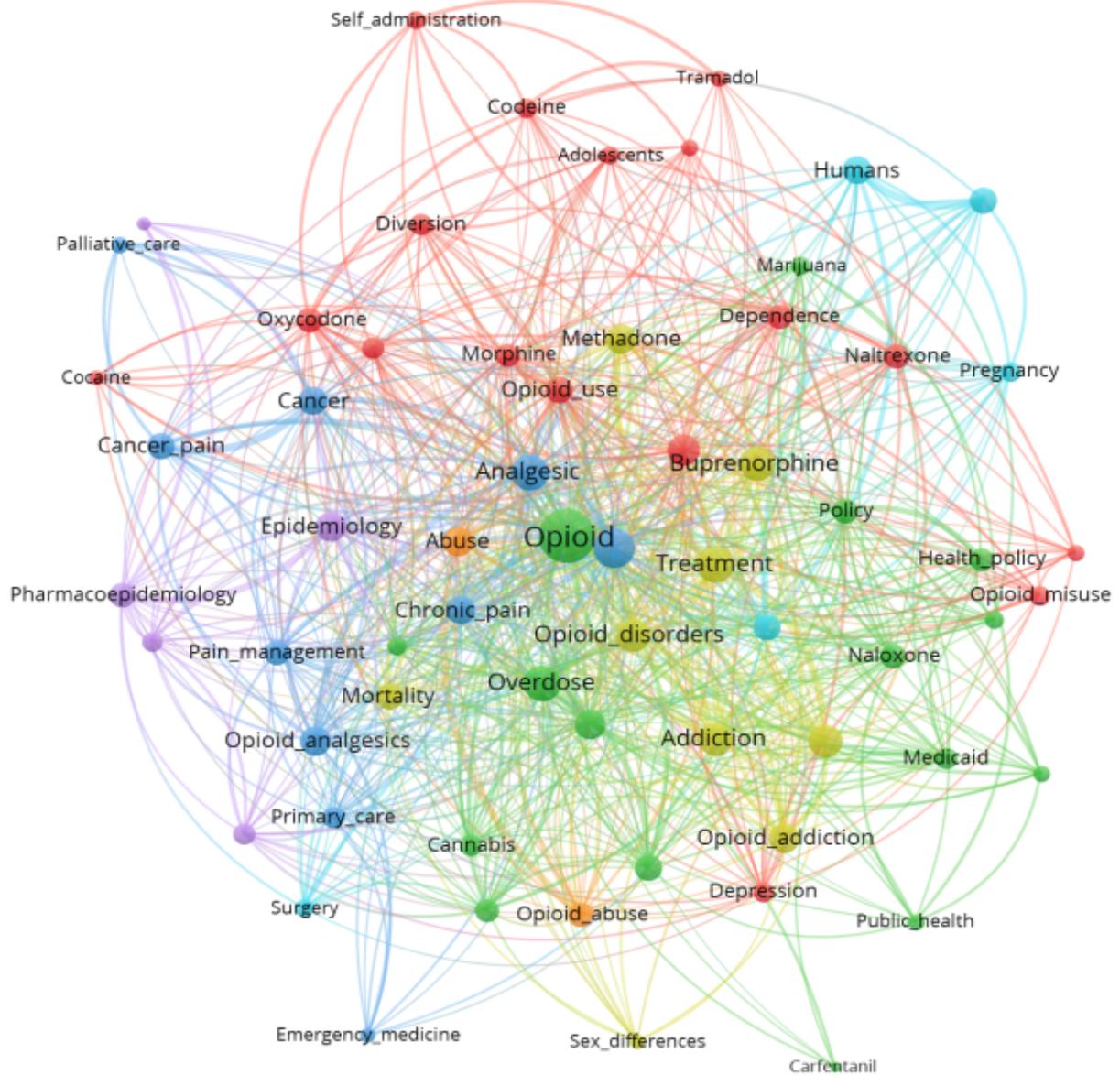

Figure 6. Network of top-64 author keywords in publications published between 2011 and 2019, tweeted by at least two accounts (considering only non-bot accounts) and mentioned in the news. The interactive network can be found at: https://s.gwdg.de/LvH6dS.

## 4.2. Hashtags

Hashtags are metadata that are often used strategically to label and describe social media posts. We analyzed hashtags in the tweets linked to opioid scholarly publications to understand how people describe those publications by using hashtags. Figure 7 shows the network of top-70 hashtags in the tweets posted by all accounts. The red color reflects the largest cluster with 27 hashtags which mostly represent palliative care and pain management. This cluster also includes research-related hashtags such as "#SCIENCE," "#OPENACCESS," "#PAINJOURNAL," and "#COCHRANE." The blue cluster is related to the opioid crisis, epidemic, and addiction. The green cluster contains hashtags like "#IDU," "#PWID," "#SUD," and "#OD," which refer to the use and injection of drugs, substance use, and substance overdose. The overlap between Figure 7 (all accounts) and Figure 8 (only non-bot accounts) is 81.3%.

Figure 7. Network of top-70 hashtags in the tweets linked to opioid scholarly publications published between 2011 and 2019 (considering all accounts). The interactive network can be found at: https://s.gwdg.de/xumAgn.

Figure 8. Network of top-64 hashtags in the tweets linked to opioid scholarly publications published between 2011 and 2019 (considering only non-bot accounts). The interactive network can be found at: https://s.gwdg.de/jK4LHd.

### 4.3. Author keywords and hashtags

We also generated the co-occurrence network of the top-35 hashtags and top-35 author keywords, including bot accounts (see Figure 9) and non-bot accounts (see Figure 10). We found a high overlap (90.06%) between the hashtags and author keywords in the two networks. Figure 9 indicates that some hashtags and author keywords in the red cluster were synonyms such as "#PAIN" and "#PAINEVIDENCE," which are associated with pain, pain management, palliative care, and cancer pain, or "#CANCER" is related to the author keywords cancer and cancer pain. The blue cluster also consists of hashtags and keywords with similar concepts, such as "#HEROIN" and "Heroin." "#OVERDOSE" and "#OD" are associated with author keywords such as overdose and opioid overdose. Other hashtags in this cluster also seem to be related to opioid overdoses, such as "#PWID" and "#PWUD," which refer to the people who inject and use opioids. In the green cluster, opioid addiction, disorders, use, and treatment are far away and less connected to hashtags like "#HCV," "#HIV," "#METHADONE," "#CANNABIS," and "#BUPRENORPHINE."

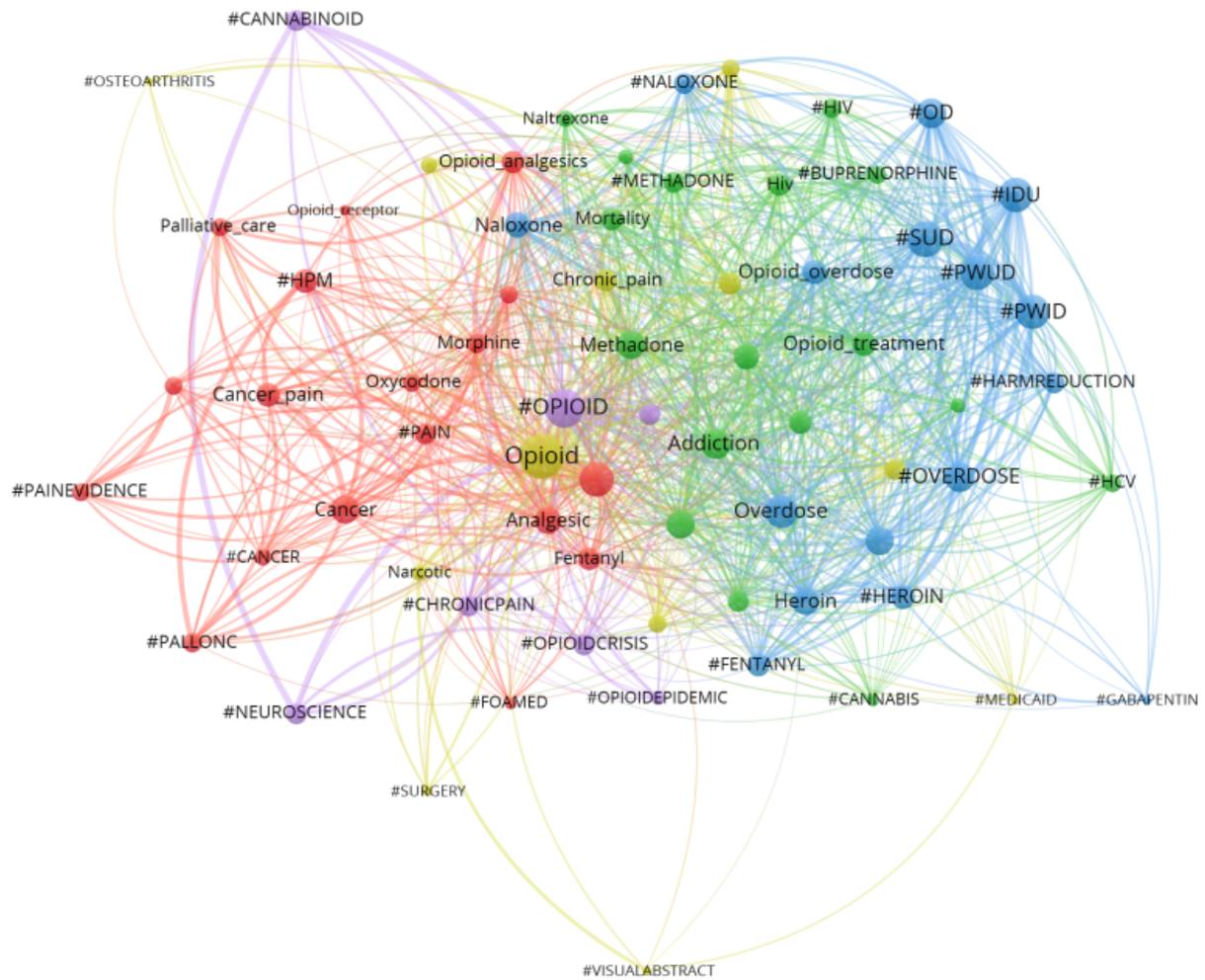

Figure 9. Network of top-35 author keywords in opioid scholarly publications published between 2011 and 2019, tweeted by at least two accounts (considering all accounts) and the top-35 hashtags in the tweets. The interactive network can be found at: https://s.gwdg.de/UPnfPm.

Figure 10. Network of top-35 author keywords in opioid scholarly publications published between 2011 and 2019, tweeted by at least two accounts (considering only non-bot accounts) and the top-35 hashtags in the tweets. The interactive network can be found at: https://s.gwdg.de/IOw481.

Table 2 shows the overlap of networks generated from the tweets posted by all accounts and non-bot accounts. There is a 95.3% overlap between the top author keywords of opioid scholarly publications tweeted by all accounts and the top author keywords of publications tweeted by only non-bot accounts. This high overlap shows that including bot accounts in the final analysis did not impact topic networks of opioid scholarly publications.

Additionally, there is a 100% overlap between top author keywords in publications tweeted by all accounts and mentioned in the news and author keywords in publications tweeted by only non-bot accounts and mentioned in the news. Table 2 indicates a high (81.3%) overlap between top hashtags of tweets by all accounts and top hashtags of tweets by only non-bot accounts. There is a 90.6% overlap between the network of top hashtags/author keywords (considering all accounts) and the network of top hashtags/author keywords (considering only non-bot accounts).

Table 2. The overlap between networks of author keywords in opioid scholarly publications posted on Twitter by bot accounts and without bot accounts

| Networks | The frequency of terms that appeared in both networks | Percentage of terms appeared in both networks |
|---|---|---|
| Overlap between Figures 3 and 4: Top author keywords of publications that were tweeted by at least two accounts | 61 | 95.3% |
| Overlap between Figures 5 and 6: Top author keywords of publications that were tweeted by at least two accounts and mentioned in the news | 64 | 100.0% |

| Overlap between Figures 7 and 8: Top hashtags of tweets | 52 | 81.3% |
| Overlap between Figures 9 and 10: Author keywords of publications that were tweeted by at least two accounts with top hashtags of tweets | 58 | 90.6% |

As evident in Table 3, there was a high overlap (84.1%) between author keywords in all opioid scholarly publications (see Figure 2) and author keywords in publications that were tweeted by at least two accounts (see Figure 3). A much lower overlap was found between author keywords in all publications and author keywords in publications that were tweeted by at least two accounts and mentioned in the news (see Figure 5). As evident in Table 3, there was a 72.5% overlap between terms in the publications that were tweeted by at least two accounts and publications that were both tweeted by at least two accounts and mentioned in the news.

Table 3. The overlap between networks of author keywords in all opioid scholarly publications and the publications tweeted that included both bot and non-bot accounts and mentioned in the news

| | Network of all publications (Figure 2) | Network of publications tweeted by at least two accounts (Figure 3) | Network of publications tweeted by at least two accounts and mentioned in the news (Figure 5) |
| --- | --- | --- | --- |
| Network of all publications (Figure 2) | 70 | 84.1% | 64.3% |
| Network of publications tweeted by at least two accounts (Figure 3) | 58 | 69 | 72.5% |
| Network of publications tweeted by at least two accounts and mentioned in the news (Figure 5) | 45 | 50 | 70 |

Table 4 shows a 78.1% overlap between the terms in all publications (see Figure 2) and publications that were tweeted by at least two accounts (see Figure 4). The overlap between the terms in all publications (see Figure 2) and publications that were tweeted by at least two accounts and mentioned in the news (see Figure 6) was 65.6%. Additionally, Table 4 shows a 79.7% overlap between terms in the publications tweeted by at least two accounts (see Figure 4), and the publications tweeted by at least two accounts and mentioned in the news (see Figure 6).

Table 4. The overlap between networks of author keywords in all opioid scholarly publications and the publications tweeted that included only non-bot accounts and mentioned in the news

| | Network of all publications (Figure 2) | Network of publications tweeted by at least two accounts (Figure 4) | Network of publications tweeted by at least two accounts and mentioned in the news (Figure 6) |
|---|---|---|---|
| Network of all publications (Figure 2) | 64 | 78.1% | 65.6% |
| Network of publications tweeted by at least two accounts (Figure 4) | 50 | 64 | 79.7% |
| Network of publications tweeted by at least two | 42 | 51 | 64 |

| accounts and mentioned in the news (Figure 6) | | | |
|---|---|---|---|

## 5. Discussion

The amount of information embedded in a message (e.g., a tweet) does not necessarily prompt users to share the message on social media (Potnis et al., 2020). This study also found that the topics in opioid scholarly publications, which are possibly more popular, important, or appealing, have received more public attention on Twitter. Public attention to specific topics might be a helpful metric to be used in certain aspects of research evaluation. For instance, funding agencies can evaluate research proposals based on whether the topics presented in the proposals have received public attention on Twitter.

We found a high overlap between all networks that were presented in the results section. An explanation for the large overlap between networks could be that publishers, authors, and Twitter users largely use paper keywords to tweet about newly published papers. It seems the most obvious thing to do, especially for publishers and Twitter users who do not understand the content of papers and couldn't come up with a better keyword list than the one given by the authors in the paper.

This study has several theoretical contributions to the research evaluation literature, which are mentioned below.

### 5.1. Communication channels for scholarly publications on Twitter

Our results indicated that there are two channels of communication in the diffusion of scientific information on Twitter: networked topics and networked hashtags. This provides evidence that people's communications about scholarly publications on social media have patterns in the form of networked topics. Patterns refer to the fact that people's interactions are not random but take place in patterned and systematic ways (Leeds-Hurwitz, 2009). These systematic patterns could be seen in the networks and clusters within networks. Each network consists of clusters of keywords that are related to each other by co-occurrence.

Another channel of communication was the network of topics and hashtags (see Figures 9 and 10 above). The networks of topics and hashtags reflect the multimodal nature of communication (Leeds-Hurwitz, 2009) on Twitter, that is both hashtags and author keywords are used by Twitter users to make sense of scholarly publications, and that both should be studied to understand how people communicate about scholarly publications on Twitter.

This may provide evidence that people understand, discuss, and remember scientific information as clusters of networked topics and hashtags. Each cluster within networks represents a community of users with similar interests discussing topics of interest. For example, in the red cluster in Figure 9, "#PAIN" and "#PAINEVIDENCE" are associated with terms such as "Pain," "Pain management," "Palliative care," and "Cancer pain."

Linking scientific information to hashtags on social media can also reflect the meta-communication function (Leeds-Hurwitz, 2009) of hashtags, in that they are used to clarify what the tweet is about. The advantage of using hashtags to transfer topics of scholarly publications on Twitter is that it allows users with similar interests to interact with each other without relying on gatekeepers like influential authors. The hashtags and topics in networks form communities and act as information organizers. Each cluster shows a sub-community of people with similar interests. People with similar interests come across each other and form virtual communities without any boundaries. If someone is interested in scholarly publications related to specific topics, they would likely meet people with similar interests. The networks do not only show connections, but they also show the possibility of people with similar interests come together.

### 5.1. Using topic networks to identify topics of interest in the public discourse on Twitter

Topic networks can be used to determine topics of interest related to scholarly publications in the public discourse on Twitter.

Publications with generic topics: Haunschild, Leydesdorff, Bornmann, et al. (2019) noted that climate change publications with more general keywords were more likely to be tweeted than those with jargon. Our results also indicated that generic topics were more noticeable in all topic networks. We found that the most tweeted topics were general keywords like "Opioid," "Pain," "Addiction," "Abuse," "Depression," "Treatment," and "Analgesics." The network of opioid publications tweeted by at least two accounts (see Figure 3 above), was also focused on similar generic topics. It was also indicated that generic hashtags were often used in the tweets linked to opioid scholarly publications, such as "#PWUD" (people who use drugs), "#PWID" (people who inject drugs), "#SUD" (substance use drugs), "#ADDICTION," "#OPIOID," "#OPIOIDCRISIS," "#OPIOIDEPIDEMIC," and "#OVERDOSE."

Besides publications with generic terms, some networks indicated public attention to specific topics related to opioids, such as policy and scientific evidence about opioids.

Policy-related topics: Publications always make recommendations such as policy recommendations relevant to government agencies, the public, and different communities. The topic networks of author keywords in publications tweeted by at least two accounts and mentioned in the news (see Figures 5 and 6 above) revealed policy-related topics. In Figure 5, "policy" is in the center of the purple cluster and linked to topics such as "Health policy," "Marijuana," "Medicaid," "Opioid prescription," and "Overdose." This result indicates the importance of policy-making regarding opioid use and prescription and its impact on health. Haunschild, Leydesdorff, Bornmann, et al. (2019) also found policy-related issues such as "Food_security," "Governance," and "Renewable_energy" in the tweeted publications about climate change. Policy-related topics may be at the focal point of public attention in other disciplines too.

Scientific evidence about opioids: We found some science-related hashtags such as "#SCIENCE," "#OPENACCESS," "#PAINJOURNAL," "#COCHRANE," "#RESEARCH," and "#STUDY." These hashtags may reflect the fact that scientists are also active on Twitter posting scholarly publications or that people attribute such hashtags to tweets to cite scientific evidence to verify the credibility of their opioid-related posts.

### 5.2. The network agenda-setting effect of scholarly publications on public attention

The current study examined which author keywords in opioid scholarly publications from 2011 to 2019 have received public attention on Twitter. We found a high overlap between the network of author keywords in all publications (Figure 2) and the networks of author keywords in the publications shared on Twitter by either bot accounts or non-bot accounts. These results show that the topics frequently discussed in scholarly publications also get much attention on Twitter. The high overlap may also represent the presence of a network agenda-setting effect of scholarly publications on public attention on Twitter (Guo, 2015). In line with the network agenda-setting model (Guo & McCombs, 2011), it can be the case that topics of scholarly publications are transferred to the public discussion on Twitter as clusters of networked topics and impact public opinion. This assumption needs to be tested in future studies using statistical tests such as Granger causality (Vargo & Guo, 2017).

The use of hashtags in tweets can also bring users with similar interests together on social media, form communities (Potnis & Tahamtan, 2021), and consequently set a public agenda on social media (Hemsley, 2019; Potnis & Tahamtan, 2021). Nevertheless, this study provides some preliminary evidence on the presence of a network agenda-setting effect on social media by the shared scholarly publications. The topics in scholarly publications that have the potential to set an agenda on social media can be said to have an impact on social media users.

### 5.3.  Effect of bots on public attention to scholarly publications on social media

We studied the impact of different actors (bot accounts versus non-bot actors) on public attention to scholarly publications. If bot accounts impact public attention to scholarly publications, they should be removed in studies that use social media data.

We investigated whether the networks generated by all accounts (bot and non-bot accounts) were different from those generated by non-bot accounts. In line with past studies (Didegah et al., 2018; Haustein et al., 2016), the results in the current research demonstrated that bot accounts were extensively involved in disseminating opioid scholarly publications on Twitter. However, because of the high overlap between the networks that included bot accounts and those with non-bot accounts, it can be said that bots do not manipulate but maybe magnify public attention to scholarly publications on Twitter. Investigating this assumption requires further studies and analyses such as multiple regression models, including larger datasets and other disciplines. In this regard, Lokot and Diakopoulos (2016) maintained that bots could be useful in automating the spread of news-by-news agencies and citizen journalists. However, their result on bots is in disagreement with parts of the literature maintaining that bots manipulate and shape public opinion and attention regarding ideological and political debates (Kollanyi et al., 2016).

### 5.4. Study limitations and future research

This study analyzed opioid scholarly publications published between 2011 and 2019 (and their corresponding tweets) that have been mentioned by at least two accounts. We did not verify if the same person owned the two accounts. However, verifying such a possibility seems to be impossible since some persons may not use their real names for their Twitter accounts. It is indeed impossible, at least with current social media platforms, to reliably tell whether the same person owns two social media accounts.

We only considered top (64 to 70) author keywords/hashtags in generating the networks and assessing the overlap between networks. Including all author keywords/hashtags may result in different overlap scores but may generate complex networks that are difficult to analyze.

The majority of the authors of scholarly publications are not on Twitter. This is a limitation that is beyond the control of the researchers in the current study. Nevertheless, the approach proposed in this study can still be applied to different domains to evaluate the impact of scholarly publications on social media.

We noted that one way to assess public attention to scholarly publications is to analyze the contents of scholarly publications shared on social media. This statement may not hold if the people sharing content on social media are the authors of opioid research publications or researchers from other fields. Therefore, future studies should find ways to classify Twitter accounts to non-academic users (representing the public) and non-public (e.g., health care professionals, journalists, organizations). In this regard, two recent studies by Mohammadi et al. (2020) and Mohammadi et al. (2018) indicated that almost half of scholarly publications in the study samples were discussed by non-academic users on Twitter and Facebook. Mohammadi et al. (2018) conducted a survey study on 1,912 Twitter users who had tweeted journal articles and indicated that half of them did not work in academia. Mohammadi et al. (2020) manually classified users who had mentioned 500 journal articles on Facebook and indicated that 58% of users were non-academics. These two studies suggested that half of the discussions on scholarly publications on social media are performed by the public. We suggest that future studies focus on methods and ways to analyze and interpret results in terms of non-academics and academics. In other words, the focus should be on topics of scholarly publications receiving more attention from non-academics compared to academics on social media.

The high overlap between networks with all accounts and networks with only non-bot accounts raises several questions that can be addressed in future studies: Are bots simply replicating and amplifying the message in tweets by humans? Does it mean that bot-generated tweet topics are subsets of human-generated tweet topics or vice-a-versa? Do bots tweet the same tweets after human-generated tweets? What is the timeline correlation between human and bot-generated tweets? Future studies should also identify how many people read the tweet by bots and non-bot accounts? The number of people who read or engage with that tweet might partially represent the magnitude of public attention.

Future research might investigate how the approach proposed in the current study works in other domains. Further research is required to confirm our result that bot accounts can impact or manipulate public attention to scholarly publications on Twitter.

## 6. Conclusions

Unlike most previous studies that have used 'the number of mentions' of scholarly publications on social media to measure research impact, the current study used topic networks to measure public attention to opioid publications. Results indicated that Twitter provides generic information about scholarly publications in the form of clusters of networked topics and hashtags. Bots are greatly involved in the distribution of scholarly publications on Twitter. However, they have a negligible impact on the networks generated from author keywords in publications. This study provided some preliminary evidence that scholarly publications might have a network agenda-setting effect on Twitter, in that the networks of topics in scholarly publications can impact public attention on Twitter.

## 7. Acknowledgment

## 8. References

Alperin, J. P., Gomez, C. J., & Haustein, S. (2019). Identifying diffusion patterns of research articles on Twitter: A case study of online engagement with open access articles. *Public Understanding of Science, 28*(1), 2-18.

Bornmann, L. (2014). Do altmetrics point to the broader impact of research? An overview of benefits and disadvantages of altmetrics. *Journal of Informetrics, 8*(4), 895-903.

Candy, B., Jones, L., Vickerstaff, V., Larkin, P. J., & Stone, P. (2018). Mu-opioid antagonists for opioid-induced bowel dysfunction in people with cancer and people receiving palliative care. *Cochrane Database of Systematic Reviews*(6). https://doi.org/10.1002/14651858.CD006332.pub3

Centers for Disease Control and Prevention. (2018a). *2018 Annual Surveillance Report of Drug-Related Risks and Outcomes — United States Surveillance Special Report*. https://www.cdc.gov/drugoverdose/pdf/pubs/2018-cdc-drug-surveillance-report.pdf

Centers for Disease Control and Prevention. (2018b). *2018 Annual Surveillance Report of Drug-Related Risks and Outcomes — United States Surveillance Special Report*. Centers for Disease Control and Prevention, U.S. Department of Health and Human Services. https://www.cdc.gov/drugoverdose/pdf/pubs/2018-cdc-drug-surveillance-report.pdf

Centers for Disease Control and Prevention. (2019a). *2019 Annual Surveillance Report of Drug-Related Risks and Outcomes — United States Surveillance Special Report*. https://www.cdc.gov/drugoverdose/pdf/pubs/2019-cdc-drug-surveillance-report.pdf

Centers for Disease Control and Prevention. (2019b). *2019 Annual Surveillance Report of Drug-Related Risks and Outcomes — United States Surveillance Special Report*. Centers for Disease Control and Prevention, U.S. Department of Health and Human Services. https://www.cdc.gov/drugoverdose/pdf/pubs/2019-cdc-drug-surveillance-report.pdf

Centers for Disease Control and Prevention. (2019c). *CDC Guideline for Prescribing Opioids for Chronic Pain*. cdc.gov/drugoverdose/prescribing/guideline.html

Chan, A. K. M., Nickson, C. P., Rudolph, J. W., Lee, A., & Joynt, G. M. (2020). Social media for rapid knowledge dissemination: early experience from the COVID-19 pandemic. *Anaesthesia*, 10.1111/anae.15057. https://doi.org/10.1111/anae.15057

Clarivate Analytics. (2018). *KeyWords Plus generation, creation, and changes*. https://support.clarivate.com/ScientificandAcademicResearch/s/article/KeyWords-Plus-generation-creation-and-changes?language=en_US

Davis, C. A., Varol, O., Ferrara, E., Flammini, A., & Menczer, F. (2016). *Botornot: A system to evaluate social bots* WWW'16 Companion, Montréal, Québec, Canada. http://dx.doi.org/10.1145/2872518.2889302a

Didegah, F., Mejlgaard, N., & Sørensen, M. P. (2018). Investigating the quality of interactions and public engagement around scientific papers on Twitter. *Journal of Informetrics, 12*(3), 960-971.

Doleman, B., Leonardi-Bee, J., Heinink, T. P., Bhattacharjee, D., Lund, J. N., & Williams, J. P. (2018). Pre-emptive and preventive opioids for postoperative pain in adults undergoing all types of surgery. *Cochrane Database of Systematic Reviews*(12). https://doi.org/10.1002/14651858.CD012624.pub2

Ferrara, E. (2020a). # covid-19 on twitter: Bots, conspiracies, and social media activism. *arXiv preprint arXiv:2004.09531*.

Ferrara, E. (2020b). What types of COVID-19 conspiracies are populated by Twitter bots? *First Monday, 25*(6). https://doi.org/https://doi.org/10.5210/fm.v25i6.10633

Guo, L. (2015). Semantic network analysis, mind mapping and visualization: a methodological exploration of the network agenda setting model. *The Power of Information Networks*, 37-51.

Guo, L., & McCombs, M. (2011). Network agenda setting: A third level of media effects. Annual conference of the International Communication Association, Boston, MA.

Haunschild, R., Leydesdorff, L., & Bornmann, L. (2019, 2-5 September 2019). *Library and Information Science papers as Topics on Twitter: A network approach to measuring public* *attention* ISSI 2019 – 17th International Conference of the International Society for Scientometrics and Informetrics, Rome, Italy.

Haunschild, R., Leydesdorff, L., & Bornmann, L. (2020). Library and information science papers discussed on Twitter: A new network-based approach for measuring public attention. *Journal of Data and Information Science, 5*(3), 5-17.

Haunschild, R., Leydesdorff, L., Bornmann, L., Hellsten, I., & Marx, W. (2019). Does the public discuss other topics on climate change than researchers? A comparison of explorative networks based on author keywords and hashtags. *Journal of Informetrics, 13*(2), 695-707.

Haunschild, R., Leydesdorff, L., Bornmann, L., Hellsten, I., & Marx, W. (2020). Corrigendum to "Does the public discuss other topics on climate change than researchers? A comparison of explorative networks based on author keywords and hashtags"[J. Informetrics 13 (2019) 695–707]. *Journal of Informetrics, 14*(1), 101020.

Haustein, S., Bowman, T. D., Holmberg, K., Tsou, A., Sugimoto, C. R., & Larivière, V. (2016). Tweets as impact indicators: Examining the implications of automated "bot" accounts on T witter. *Journal of the Association for Information Science and Technology, 67*(1), 232-238.

Havakhor, T., Soror, A. A., & Sabherwal, R. (2018). Diffusion of knowledge in social media networks: effects of reputation mechanisms and distribution of knowledge roles. *Information systems journal, 28*(1), 104-141.

Hegelich, S., & Janetzko, D. (2016). Are social bots on Twitter political actors? Empirical evidence from a Ukrainian social botnet. Tenth International AAAI Conference on Web and Social Media,

Hellsten, I., & Leydesdorff, L. (2020). Automated analysis of actor–topic networks on twitter: New approaches to the analysis of socio-semantic networks. *Journal of the Association for Information Science and Technology, 71*(1), 3-15.

Hemsley, J. (2019). Followers Retweet! The Influence of Middle-Level Gatekeepers on the Spread of Political Information on Twitter. *Policy & Internet, 11*(3), 280-304.

Kamada, T., & Kawai, S. (1989). An algorithm for drawing general undirected graphs. *Information processing letters, 31*(1), 7-15.

Kearney, M. W. (2019). *tweetbotornot*. Retrieved January 28 from https://github.com/mkearney/tweetbotornot

Kollanyi, B., Howard, P. N., & Woolley, S. C. (2016). Bots and automation over Twitter during the first US presidential debate. *Comprop data memo, 1*, 1-4.

Korda, H., & Itani, Z. (2013). Harnessing social media for health promotion and behavior change. *Health promotion practice, 14*(1), 15-23.

[Record #47 is using a reference type undefined in this output style.]

Leidl, R. (2019). Social media, bots and research performance. *European Journal of Public Health, 29*(1), 1-1. https://doi.org/10.1093/eurpub/cky123

Lokot, T., & Diakopoulos, N. (2016). News Bots: Automating news and information dissemination on Twitter. *Digital Journalism, 4*(6), 682-699.

Moe-Byrne, T., Brown, J. V. E., & McGuire, W. (2018). Naloxone for opioid-exposed newborn infants. *Cochrane Database of Systematic Reviews*(10). https://doi.org/10.1002/14651858.CD003483.pub3

Mohammadi, E., Barahmand, N., & Thelwall, M. (2020). Who shares health and medical scholarly articles on Facebook? *Learned Publishing, 33*(2), 111-118.

Mohammadi, E., Thelwall, M., Kwasny, M., & Holmes, K. L. (2018). Academic information on Twitter: A user survey. *PloS one, 13*(5), e0197265.

Müller, K., Wickham, H., James, D. A., & Falcon, S. (2017). *RSQLite:'SQLite'Interface for R. R package version 1.1-2*. https://rsqlite.r-dbi.org/

Murphy, G., & Salomone, S. (2013). Using social media to facilitate knowledge transfer in complex engineering environments: a primer for educators. *European Journal of Engineering Education, 38*(1), 70-84.

National Institute on Drug Abuse. (2019). *What are prescription opioids?* https://www.drugabuse.gov/publications/drugfacts/prescription-opioids

Patel, V., Haunschild, R., Bornmann, L., & Garas, G. (2020). A call for governments to pause Twitter censorship: a cross-sectional study using Twitter data as social-spatial sensors of COVID-19/SARS-CoV-2 research diffusion. *medRxiv*. https://doi.org/https://doi.org/10.1101/2020.05.27.20114983

Potnis, D., & Tahamtan, I. (2021). Hashtags for gatekeeping of information on social media. *Journal of the Association for Information Science and Technology*, 1-13. https://doi.org/https://doi.org/10.1002/asi.24467

R Core Team. (2019). *R: A language and environment for statistical computing.* In (Version 3.6.0) R Foundation for Statistical Computing. https://www.r-project.org/

R Special Interest Group on Databases (R-SIG-DB), Wickham, H., & Müller, K. (2018). https://dbi.r-dbi.org/

Rubinstein, A. L., & Carpenter, D. M. (2017). Association between commonly prescribed opioids and androgen deficiency in men: a retrospective cohort analysis. *Pain Medicine, 18*(4), 637-644.

Rudd, R. A. (2016). *Increases in drug and opioid-involved overdose deaths—United States, 2010–2015* (MMWR. Morbidity and Mortality weekly Report, Issue. http://dx.doi.org/10.15585/mmwr.mm655051e1

Rudd, R. A., Aleshire, N., Zibbell, J. E., & Matthew Gladden, R. (2016). Increases in drug and opioid overdose deaths—United States, 2000–2014. *American Journal of Transplantation, 16*(4), 1323-1327.

Smith, L. A., Burns, E., & Cuthbert, A. (2018). Parenteral opioids for maternal pain management in labour. *Cochrane Database of Systematic Reviews*(6). https://doi.org/10.1002/14651858.CD007396.pub3

Sugimoto, C. R., Work, S., Larivière, V., & Haustein, S. (2017). Scholarly use of social media and altmetrics: A review of the literature. *Journal of the Association for Information Science and Technology, 68*(9), 2037-2062.

Tahamtan, I., & Bornmann, L. (2020). Altmetrics and societal impact measurements: Match or mismatch? A literature review. *El profesional de la información (EPI), 29*(1). https://doi.org/https://doi.org/10.3145/epi.2020.ene.02

The White House. (2017). *Ending America's Opioid Crisis*. https://www.whitehouse.gov/opioids/

Vargo, C. J., & Guo, L. (2017). Networks, big data, and intermedia agenda setting: An analysis of traditional, partisan, and emerging online US news. *Journalism & mass communication quarterly, 94*(4), 1031-1055.

## 9. Appendices

### 9.1. Appendix A

We performed our search including the final list of search terms in WoS in three steps (#1, #2, and #3), as follows:

**#1:** We assume that English publications between 2011 and 2019 that contain "Opiate* OR Opioid*" in their titles are related to opioid research:

- (TI=(Opiate* OR Opioid*)) AND LANGUAGE: (English) AND DOCUMENT TYPES: (Article OR Review) Indexes=SCI-EXPANDED, SSCI, A&HCI Timespan=2011-2019

**#2:** We then executed the following search in WoS to retrieve additional English publications which also appeared between 2011 to 2019. This search strategy retrieves scholarly publications that contain opioid-related terms in their titles (TI field) and also contain "Opiate* OR Opioid*" in their title, abstract, or author keywords (TS[1]). The asterisk sign (*) at the end of the search terms

---

1 TS or the topic field searches in title, abstract, author keywords, and KeyWords Plus. KeyWords Plus are generated by the database provider from the titles of cited documents. "The data in KeyWords Plus are words or phrases that frequently appear in the titles of an article's references but do not appear in the title of the article itself" Clarivate Analytics. (2018). *KeyWords Plus generation, creation, and changes*. https://support.clarivate.com/ScientificandAcademicResearch/s/article/KeyWords-Plus-generation-creation-and-changes?language=en_US.

was used to find variations of the search terms (e.g., "Opioid*" finds both "Opioid" and "Opioids"):

- (TI=(Narcotic* OR Morphine OR Heroin OR Suboxone OR Subutex OR Kadian OR Avinza OR Diamorphine OR Fentanyl OR Remifentanil OR Alfentanil OR Meperidine OR Pethidine OR Tramadol OR Ketobemidone OR Hydrocodone OR Vicodin OR Hydromorphone OR Methadone OR Oxycodone OR OxyContin OR Percocet OR Oxymorphone OR Opana OR Tapentadol OR Codeine OR Buprenorphine OR Butrans OR Belbuca OR Propoxyphene) AND TS=(Opiate* OR Opioid*)) AND LANGUAGE: (English) AND DOCUMENT TYPES: (Article OR Review) Indexes=SCI-EXPANDED, SSCI, A&HCI Timespan=2011-2019

**#3**: The above two searches provided us with two sets of opioid scholarly publications. However, there is some overlap between the publications retrieved in steps #1 and #2. Therefore, we combined the two sets using the OR operator (#1 OR #2). This approach removed duplicated publications and resulted in 16,889 publications.

### 9.2. Appendix B

The downloaded data was stored in a local SQLite database file using the R package "RSQLite" (Müller et al., 2017). Functions from the R package "DBI" (R Special Interest Group on Databases (R-SIG-DB) et al., 2018) were used for sending database queries. The R package "tweetbotornot" designed by Kearney (2019) was used to detect bots.

### 9.3. Appendix C

We used a threshold to exclude the tweets that were not available. Since the tweets' texts and meta information of available tweets are usually longer than 90 characters, we used this as a threshold to filter for available or unavailable tweets. We found 165,660 available tweets and 7,527 unavailable tweets. Also, some tweets were not downloaded because of Twitter's internal errors (internal and overcapacity errors). However, this applied to only three tweets, two with the "overcapacity" error and one with the "internal" error. Additionally, 1,516 tweets, from accounts suspended by Twitter, were not included in our analysis.